\documentclass{aa}
\usepackage{graphicx}
\usepackage{txfonts}
\usepackage{amsmath}
\usepackage{soul}
\usepackage{cancel}

\usepackage{pdfpages}
\begin{document} 

     \title{Testing MURaM and MPS-ATLAS against the quiet solar spectrum}
  
   \author{Veronika~Witzke\inst{1}\fnmsep\thanks{e-mail: witzke@mps.mpg.de}
          \and
          Alexander~I.~Shapiro\inst{1} \and
          Nadiia~M.~Kostogryz\inst{1} \and
          Lucien Mauviard \inst{1,2} \and
          Tanayveer~S.~Bhatia\inst{1}\and  
          Robert~Cameron\inst{1}\and
          Laurent~Gizon\inst{1,3,4} \and
          Damien~Przybylski\inst{1}\and
          Sami~K.~Solanki\inst{1}\and
          Yvonne~C.~Unruh\inst{5}\and 
          Li~Yue\inst{1,6}
          }

   \institute{Max Planck Institute for Solar System Research, Justus-von-Liebig-Weg 3, 37077 G\"ottingen, Germany\\ 
\and
Institut Supérieur de l’Aéronautique et de l’Espace (ISAE-SUPAERO), 31400 Toulouse, France\\
\and
Institut f\"ur Astrophysik, Georg-August-Universit\"at G\"ottingen, Friedrich-Hund-Platz 1, 37077 G\"ottingen, Germany\\
\and
Center for Space Science, NYUAD Institute, New York University Abu Dhabi, PO Box 129188, Abu Dhabi, UAE\\
\and
Department of Physics, Imperial College London, London SW7 2AZ, UK\\
\and
Changchun Institute of Optics, Fine Mechanics and Physics, Chinese Academy of Sciences, China\\
}
   \date{---}

  \abstract
   {Three-dimensional (3D) radiative magnetohydrodynamics (MHD) simulations are the only way to model stellar atmospheres without any ad hoc parameterisations. 
   Several 3D radiative MHD codes have achieved good quantitative agreement with observables for our Sun.}
   {We aim to validate the most up-to-date version of the MURaM code against well established quiet Sun measurements, in particular spatially averaged measurements that are relevant for stellar studies. This validation  extends the number of solar observables that MURaM can reproduce with high precision. It  is also an essential condition for using MURaM to accurately calculate spectra of other cool stars.}
   {We simulate the solar upper convection zone and photosphere harbouring a  small-scale-dynamo. Using time series of 3D snapshots we calculate the spectral irradiance, limb darkening and selected spectral lines, which we compare to observations. }
   {The computed observables agree well with the observations, in particular the limb darkening of the quiet Sun is reproduced remarkably well.}
   {}


\maketitle

\section{Introduction}
The conditions and physical processes in the solar photosphere are complex, as not only the radiative transfer in the photosphere but also the convective motion in the layer beneath it, the convection zone, and the magnetic field plays a role. Accordingly,  the solar photosphere needs to be described by means of radiative magnetohydrodynamics (MHD). 
A milestone was reached when realistic 3D hydrodynamic (HD, i.e. without magnetic fields) simulations of the lower solar atmosphere became numerically feasible \citep{nordlund_1982} and allowed reproducing the transfer of energy in stellar atmospheres without  any additional ad hoc parameterisations \citep[e.g., mixing lengths approximation, see][]{1958ZA.....46..108B}.  %

Due to the success of solar 3D HD simulations, they were extended to other cool stars, creating grids of stellar models that cover the spectral classes F--M  \citep{Magic_2013A&A}. Such HD simulations were used for numerous applications, e.g. for modelling stellar limb darkening \citep{Magic2015A&A_paperIV} and stellar spectra \citep{Chiavassa2018}.
At the same time, MHD simulations were introduced to consider various magnetic phenomena in the solar near-surface layers, for example, network and facular regions \citep{NordlundandStein1990, Vogler_Sch_2005}, pores (Cameron et al. 2007), and sunspots (Rempel et al. 2009).

In the last decade, 3D MHD simulations of the solar atmosphere were further developed and have achieved extraordinary agreement with detailed solar observations. Therefore, once more, the concept was expanded to stars other than the Sun. For example, \citet{Beeck2015A&Athird}  and \citet{Salhab_2018} calculated faculae on stars of various effective temperatures.

Until recently, quiet solar regions, although they are covered by small-scale magnetic fields of mixed polarity (which are sometimes called turbulent magnetic fields)
even in their quietest state
\citep[][]{Khomenko2003A&A,sveta2004, Trujillo2004Natur,Lites2004ApJ,Stenflo2013},  were modelled assuming the non-magnetic HD simulations. This changed with the effort to model the solar quiet regions using the state-of-the-art code \mbox{MURaM} and including a small-scale dynamo (SSD) driven by near-surface convection \citep{Voegler_Schuessler_2007A&A, Schuessler_voegler2008, rempel_2014}. 
With the new version of the MURaM code,  which includes an up-dated equation of state (EOS) and opacity table, we have started to calculate an extensive grid of stellar 3D model atmosphere, which also account for the action of the SSD.

Previous studies focused on spatially resolved solar observations, showing good agreement with MURaM simulations \citep[e.~g.,][]{Schuessler_et_al2003, Shelyag_et_al2004, Keller_et_al2004, Hirzberger_2010, Riethmueller_2014}. Moreover, using averaged quantities, it has been shown that facular contrasts are sufficiently well modelled to accurately reproduce TSI variations \citep{Yeo_et_al2017}. However,  careful comparisons with the quiet Sun center-to-limb variations (CLVs), solar spectral irradiance (SSI) spectrum and line bisectors have not so far been carried out.  Since these are necessary to validate the accuracy of the 3D model atmospheres for high precision stellar studies, we take on the perhaps overdue task of confronting MURaM with observations of the quiet Sun, concentrating on spatially averaged data.

The paper is structured as follow: In Section~\ref{Sec:model} we give a brief description of the codes used and review the most recent improvements of the MURaM code. Subsequently, we compare the calculated low resolution spectra and a few  high-resolution spectral lines to observations in Section~\ref{Sec:comparison} and present our conclusions in Section~\ref{Sec:conclusions}.


\section{MURaM 3D cubes and MPS-ATLAS radiative transfer}
\label{Sec:model}

To model the dynamics and energy transfer in the upper convective zone and photosphere the MURaM code takes the box-in-a-star approach. For that, the conservative MHD equations for a compressible, partially ionised plasma are solved numerically in a Cartesian box spanning the upper convection zone, the solar surface and  the photosphere. Partial ionisation is crucial for the convective energy transfer in the upper convective zone and for the transfer of radiative energy which takes over from convection as the dominant energy transport mechanism in the solar photosphere.  
The radiative transfer (RT) in the MURaM code is taken care of by a multi-group scheme \citep{nordlund_1982} with short characteristics  \citep[for a detailed description see][and references therein]{Voegler_2004}. More detailed descriptions of the governing equations and numerical implementation are given in \citet{Voegler_2004, rempel_2014,  Przybylski_2022,Bhatia_2022}.

Since the development of the MURaM code \citep{Voegler_2004, Vogler_Sch_2005} for simulations of the solar convection zone and photosphere, much effort was put into further advancement of the code to achieve more accurate calculations along with a wider range of applications. For example, boundary conditions were implemented  that take into account the effect of the deeper convection layers. This allows achieving a good agreement between the observed and modelled turbulent magnetic flux in the photosphere  \citep{rempel_2014}.  
The equation of state (EOS) look-up-table, which in the new MURaM setup are generated by the  FreeEOS code \citep{Irwin_freeeos_2012}, together with the opacity table were updated to the most accurate element composition \citep{Asplund_2009}, hereafter referred to as the Asplund composition. Furthermore, the radiative treatment was upgraded to consider 12 multi-group  bins with similar thresholds as in \citet{Magic_2013A&A}
Finally, the implementation was modified to reduce the diffusive terms as much as possible \citep{rempel_2014}.

 We use the same box size, grid resolution, and boundary conditions as in \citet{Witzke_2022_1d_3d}. In particular, the entropy inflow and pressure at the bottom are chosen such that an effective temperature of 5787 K is sustained. The simulation domain reaches 4~Mm below the optical surface into the convection zone and 1~Mm above (which covers the photosphere up to the temperature minimum) with a grid resolution of (512x512x500) grid points. 
We run 4 independent sets, each of at least one hour solar time and save cubes each 90 seconds, resulting in 160 cubes.  We have tested that this is sufficient to average out temporal fluctuations of all observables we synthesise.

To synthesize emergent spectra from the 3D simulations we apply the so called 1.5D approach, that is we calculate the radiative transfer (RT) along many mutually parallel rays passing through a 3D cube. 
Due to the large number of RT calculations  necessary for this approach, it is computationally  expensive, but is feasible using the recently developed MPS-ATLAS code \citep{mps-atlas_2021}.
MPS-ATLAS pre-tabulates opacity tables (using Kurucz's original line lists \footnote{http://wwwuser.oats.inaf.it/castelli/sources/dfsynthe.html}) either with a high resolving power allowing for separate spectral line calculations, or using the Opacity Distribution Functions (ODF) approach on a low resolution wavelength grid. We note that the opacity tables used in the MURaM code were also calculated using the MPS-ATLAS code. 
For the high-resolution spectral line synthesis we account for the Doppler effect using the line-of-sight component of the velocity  from the 3D cubes  (for a more detailed description of the Doppler shift implementation see Mauviard, in prep.). This allows us to compute both low spectral resolution emergent spectra  and profiles of individual spectral lines.

 We consider different viewing-angles, \mbox{$\rm \mu = cos(\theta)$}, where $\theta$ is the angle between the observer’s direction and the position vector defined with respect to the centre of the star. For that, we rotate the 3D cube around a pivot axis at the optical surface using the corresponding $\theta$ and compute the temperature, pressure, density and velocity along the line-of-sight from the cube. For the so obtained 3D cube, we interpolate to keep the vertical resolution of 10 km between the grid points as in the original cube. The RT calculations for each viewing-angle are then performed using the corresponding rotated cube. 
 The resulting intensity along any particular direction is obtained by combining the intensities along parallel rays, which sample the whole simulation cube. Subsequently, we take the average over all 160 cubes.  For the solar 
 irradiance and center-to-limb variations, we computed ten viewing-angles from disk-center ($\mu = 1.0$) in steps of 0.1 towards the limb. The high-resolution spectral line calculations are performed only for disk-center.

\section{Comparison to observations}
\label{Sec:comparison}

We chose three different observed quantities, the spectral irradiance (Sect.~\ref{sub:irradiance}), limb darkening (Sect.~\ref{sub:ldcomp}), as well as profiles and bisectors of spectral lines (Sect.~\ref{sub:spectral}) to test how accurately the MURaM code models the spatially averaged  solar photosphere. The vertical temperature structure of the MURaM 3D cubes can be assessed when comparing the modelled irradiance and its limb darkening to observations. In particular, the limb darkening behaviour is very sensitive to temperature in all photospheric layers. At the same time, line bisectors provide a standard test of the convective velocities \citep{Asplundetal2000} and temperature inhomogeneities produced by MURaM.

\subsection{Solar spectral irradiance}\label{sub:irradiance}

\begin{figure*}
 \sidecaption
   {\includegraphics[width=12cm]{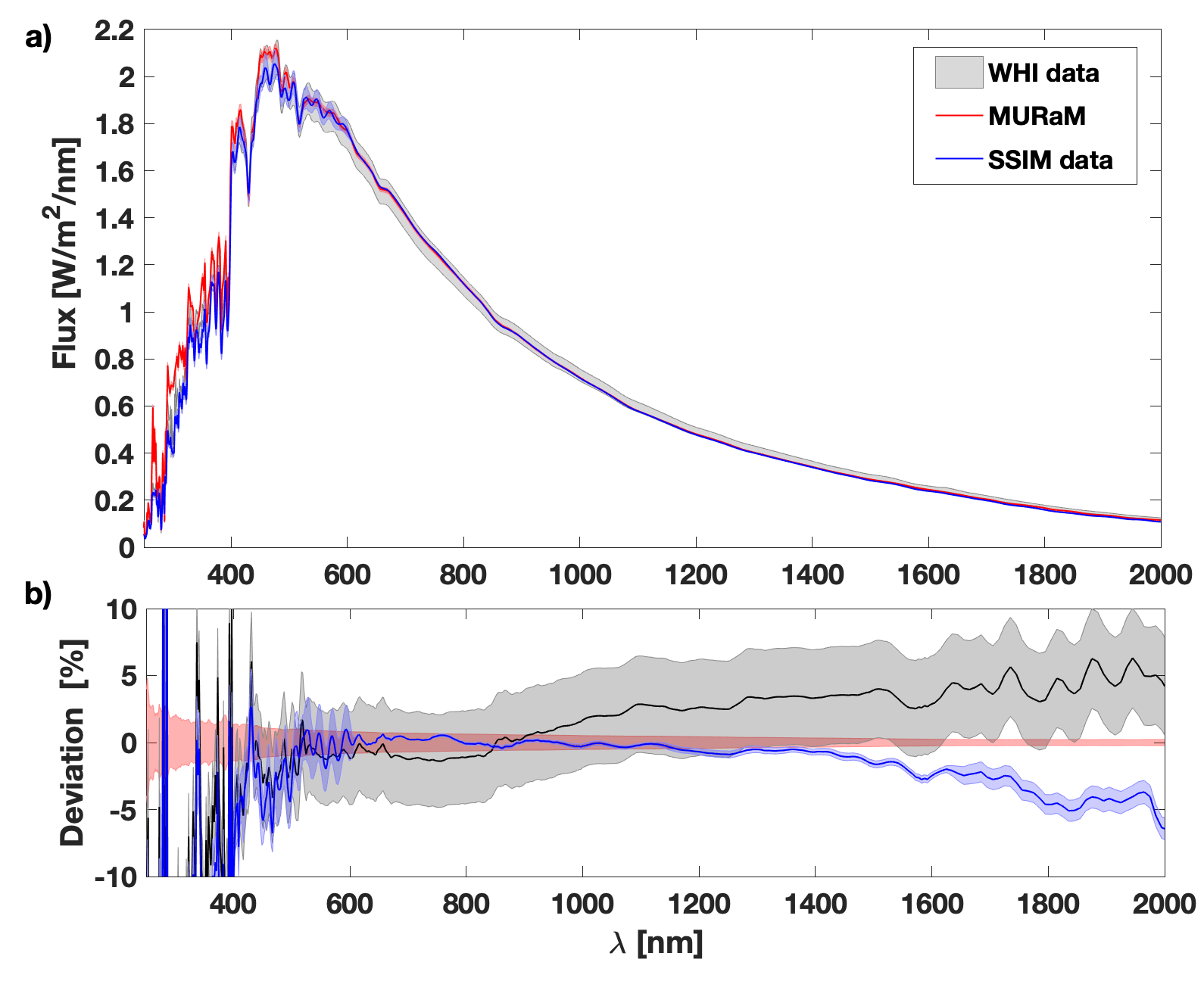}}
  \caption{Comparison of the  calculated and measured solar irradiance spectrum.
    The grey shaded area indicates the measurement error of the SIRS WHI (Solar Irradiance Reference Spectra for the 2008 Whole
Heliosphere Interval) \citep[see,][for a detailed description]{Woods_2009_WHI}. The light blue shaded area indicates one standard deviation of the 145  measurements taken in 2022 by  Solar Spectral Irradiance Monitor (SSIM)
onboard Fengyun-3E Satellite (FY-3E). The averaged solar irradiance spectrum computed from MURaM cubes is represented by the  red line, where the light red shaded area indicates one standard deviation in the temporal fluctuations.}
  \label{fig:muram_versus_observations}
\end{figure*}

For the spectral irradiance we compute emergent spectra at one AU from the Sun with a low spectral resolution using the ODF approach and the  `little' wavelength grid  from \citet{Castelli_2005_DFSYNTHE} in the interval from 250nm to 2000nm. 
In order to calculate the disk-integrated flux and also the limb darkening, we consider 10 viewing-angles starting at the disk center ($\mu = 1.0$) and proceeding in 0.1 steps towards the limb until $\mu = 0.1$. After obtaining all viewing-angles for one cube, we calculate for each viewing-angle the average over the cube and subsequently the disk-integrated flux of the cube. This procedure is then repeated for every cube   and the average over all cubes  is obtained along with the standard deviation (shown in Fig.~\ref{fig:muram_versus_observations} in red shaded band).  

We compare our calculations to Solar Irradiance Reference Spectra (SIRS) for the 2008 Whole Heliosphere Interval \citep[WHI;][]{Woods_2009_WHI}. We also considered irradiance measurements obtained by the Solar Spectral Irradiance Monitor (SSIM) onboard the Fengyun-3E Satellite (FY-3E) launched in July 2021.  SSIM measures irradiance from 165 nm to 2400 nm with a spectral resolution from 1 nm (in the UV and visible) to 8 nm (in the infrared). The measurements are performed with daily cadence and we averaged all the data from 2022. 
For a better comparison, the SSIM data and the  calculated fluxes are smoothed out using a trapezoidal kernel (Harder, private comm.) to match the WHI resolution in wavelength regions where the ODF resolution is higher than that of the WHI. We kept the original resolution otherwise.

Fig.~\ref{fig:muram_versus_observations} shows the solar irradiance calculated from the \mbox{MURaM} cubes together with the WHI and SSIM measurement.
One can see that shortwards of 350 nm our calculations overestimate the observed flux. This is a well-known problem common to all codes used for irradiance modelling \citep[see, e.g.][and references therein]{Shapiro2010, Criscuoli2019, mps-atlas_2021}. It is thought to be caused by the lack of weak atomic and molecular lines in available line lists. 
In other spectral regions our irradiance calculations are very close to the observed values and remain within the measurements uncertainties.
In particular, they remain within the WHI uncertainties  between 350~nm and 1400~nm. This demonstrates that MURaM properly reproduce the vertical temperature structure at the most of the very lowest photosphere and, especially, the transfer of energy by convective overshoot and radiation in the lower solar photosphere.
Interestingly, SSIM data deviate from SIRS by up to 10\% longward of 1600 nm with our calculations lying  in between the two measured spectra. We note that the absolute measurements of the infrared irradiance are particularly challenging and there are significant deviations between different datasets \citep[][]{IR1, IR2, IR3}.

To further understand which modifications in the MURaM simulations lead to a better match with the observations we compared spectral irradiance calculated from 3D cubes calculated using  different MURaM versions and also illustrate the effect of inconsistent abundances. 
In this comparison (shown in Figures~\ref{appfig:hd_ssd_irradiance} - \ref{appfig:muram_versus_observations} ) we  include  spectral irradiance calculations using the following different MURaM cubes: 3D cubes calculated for this study using the most up-dated MURaM version with 12 multi-group bins in the RT calculations (hereafter MULTI12). Second, pure hydrodynamic MURaM cubes with the most up-dated MURaM version, but with the same setup as our SSD calculation presented here (hereafter Hydro12). Third, 3D cubes obtained using the newest MURaM setup but with only four multi-group bins in the RT (hereafter MULTI4). Fourth, 3D cubes generated using the older MURaM version by \citet{Beeck2013A&Afirst} (hereafter Beeck2013). The spectral irradiance calculations for the fourth set of cubes was performed by \citet{Norris_2017A&A}. 

The comparison of spectral irradiance shows that differences between using MULTI4 cubes and MULTI12 cubes are significantly smaller than the difference between spectral irradiance calculations using Beeck2013 and MULTI12. Thus, the largest improvement in the spectral irradiance comes from the new EOS and opacity tables as well as potentially from the new diffusive terms in MURaM. While there are noticeable differences between Hydro12 cubes and the SSD cubes from MULTI12, SSD mainly affects velocities, whose effect will become visible in individual spectral line calculations. For a more detailed discussion see Appendix~\ref{app:irradiance}.

\subsection{Centre-to-limb variations of solar intensity} \label{sub:ldcomp}

In the next step, we compare the limb darkening computed with MURaM and post-processed with MPS-ATLAS to the measurements performed by \citet{PSW77} and \citet{NL94} at the National Solar Observatory/Kitt Peak with the large vertical spectrograph at the McMath Solar Telescope. \citet{PSW77} performed measurements at 50 continuum wavelengths in the range from 740~nm to 2402~nm, while \citet{NL94} observed at 30 continuum wavelengths in the range from 303~nm to 1099~nm. 

\begin{figure}
  \centering 
     {\includegraphics[width=\linewidth]{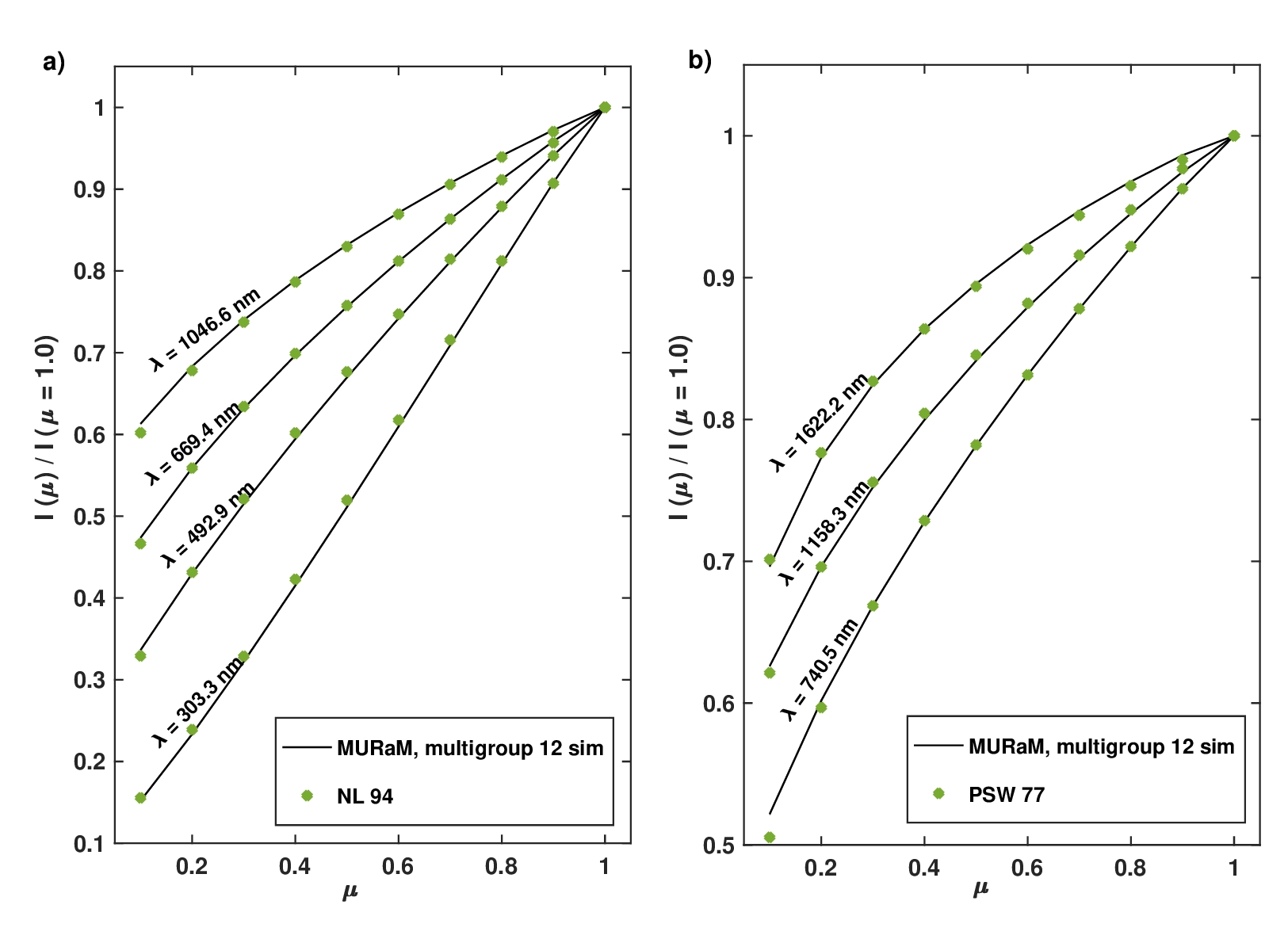}} \\
     {\includegraphics[width=\linewidth]{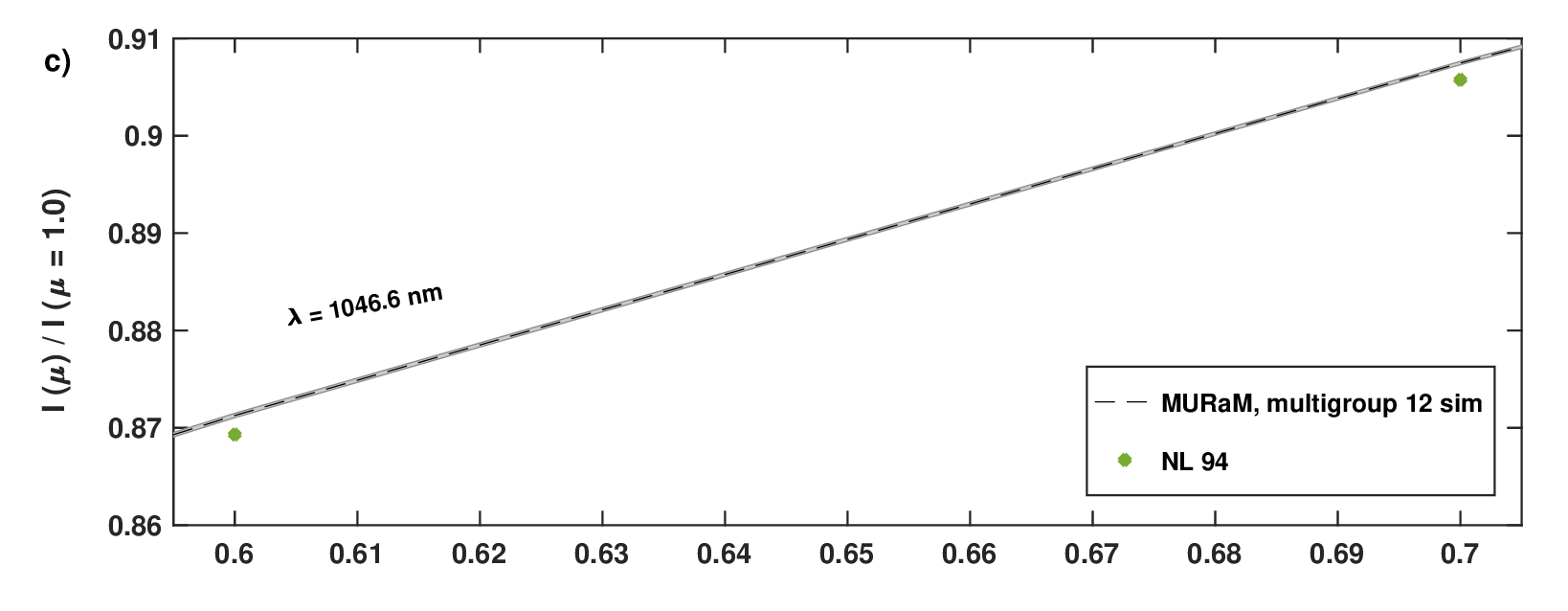}}
  \caption{Limb darkening obtained from 3D MURaM cubes (black lines) and from solar measurements  (green dots) at different wavelengths: a) measurements from NL94 \citep{NL94}, b) measurements from PSW77 \citep{PSW77}. We plotted a three sigma error of the mean for the calculated CLV from MURaM as grey shaded area, which an be seen in the zoom plot (c).}
  \label{fig:02-ld-comparison}
\end{figure}

Both studies fitted the derived limb darkening curves by fifth-order polynomials and presented the obtained coefficients. We note, that for the fitting procedure the intensity at the limb \citep[$\mu<0.12$ for ][]{NL94}  and \citep[$\mu<0.2$ for][]{PSW77} were excluded which implies a somewhat lower accuracy of the measurements at $\mu=0.1$, particularly for the values found by \citet{PSW77}.
Fig.~\ref{fig:02-ld-comparison} shows the comparison of limb darkening curves, $\rm I (\mu) / I(1.0)$, calculated from these polynomial coefficients to our computations with MURaM and MPS-ATLAS. The comparison is performed in a broad spectral range  from the UV (303.3~nm) to the wavelength corresponding roughly to the H- opacity minimum (1622.2~nm), where the disk center intensity originates in the deepest layers of solar photosphere. 
The MURaM and MPS-ATLAS calculations are in  remarkable agreement with the measured values. Deviations are mainly noticeable very close to the limb at $\mu$ value of $0.1$ at which the plotted symbols (red) do not represent direct measurements, but  rather correspond to extrapolation from measurements  at larger $\mu$ values. Further, minor deviations are also visible at $\mu=0.8$ and $\mu=0.9$ for the wavelengths 1158.3~nm and 1622.2~nm. The deviations vary with wavelength and thus might be explained by noisy observational data at these wavelengths.

To pinpoint which improvement in the MURaM simulation result in this remarkable match, we compared the present results with the CLVs from calculations by \citet{Norris_2017A&A}. Fig.~\ref{appfig:02-ld-comparison} shows that the accurate match of the CLVs comes from the increase in the number of spectral bins used for multi-grouping (for a detailed discussion see Appendix~\ref{app:clv}).


\subsection{Spectral lines and their bisectors}\label{sub:spectral}

\begin{table}
\setlength\tabcolsep{5pt}
\renewcommand{\arraystretch}{1.5}
\caption{Selected spectral lines and their atomic parameters. We list the wavelength, $\lambda_0$ in vacuum, the excitation energy of the lower leverl, $\chi_1$, the  logarithm of the oscillator strength times the multiplicity of the lower level, log $gf$, the effective Land\'e factor, $g_{eff}$, and the radiative damping constant, $\gamma_{rad}$.}
\label{table:line_parameter+}      
\centering                                     
\begin{tabular}{|| c || c  c c c c ||}

Line &  $\lambda_0$ [nm] &  $\chi_1$ [eV] & log $gf$ & $g_{eff}$ & log $\gamma_{rad}$ \\   
\hline              
Fe~I~624  & 624.2367 & 2.220 & -3.233 & 1.0 &  6.880  \\
Fe~I~627  & 627.3004  & 3.329 & -2.703 & 1.5 & 8.230 \\
Fe~I~680  & 680.6141 & 4.575 & -1.813 & 1.15 & 7.850  \\
Fe~II~771 & 771.3845  & 3.903 & -2.47 & 1.4 & 8.615 \\

\end{tabular}
\end{table}

Having tested low-resolution spectra emerging from MURaM cubes against observations, here we focus on profiles of individual spectral lines as well as their bisectors. While low  spectral resolution spectra are mainly defined by the vertical temperature structure of the cubes, the detailed shape of spectral lines and especially their bisectors are determined by the convective velocities and temperature contrasts and, thus, allow us to test if MURaM can accurately reproduce them.
For that we consider four iron spectral lines which were also used by \citet{Asplundetal2000} 
and which are only slightly blended by neighbouring lines (listed in Table~\ref{table:line_parameter+}):  the line Fe\,I (624.2~nm), the line Fe\,I (627.3~nm), the line Fe\,I (680.6~nm) and  the line Fe\,II (771.3~nm). We chose two weak and two rather strong lines with significantly different excitation energies of the lower level, $\chi_1$.  Since $\chi_1$ defines the temperature sensitivity of the line it is expected that the bisectors of the lines will show different shapes \citep[e.g.,][]{Dravins_et_al1981, Dravins_1986A&A}.

\begin{figure}
  \centering 
   {\includegraphics[width=\linewidth]{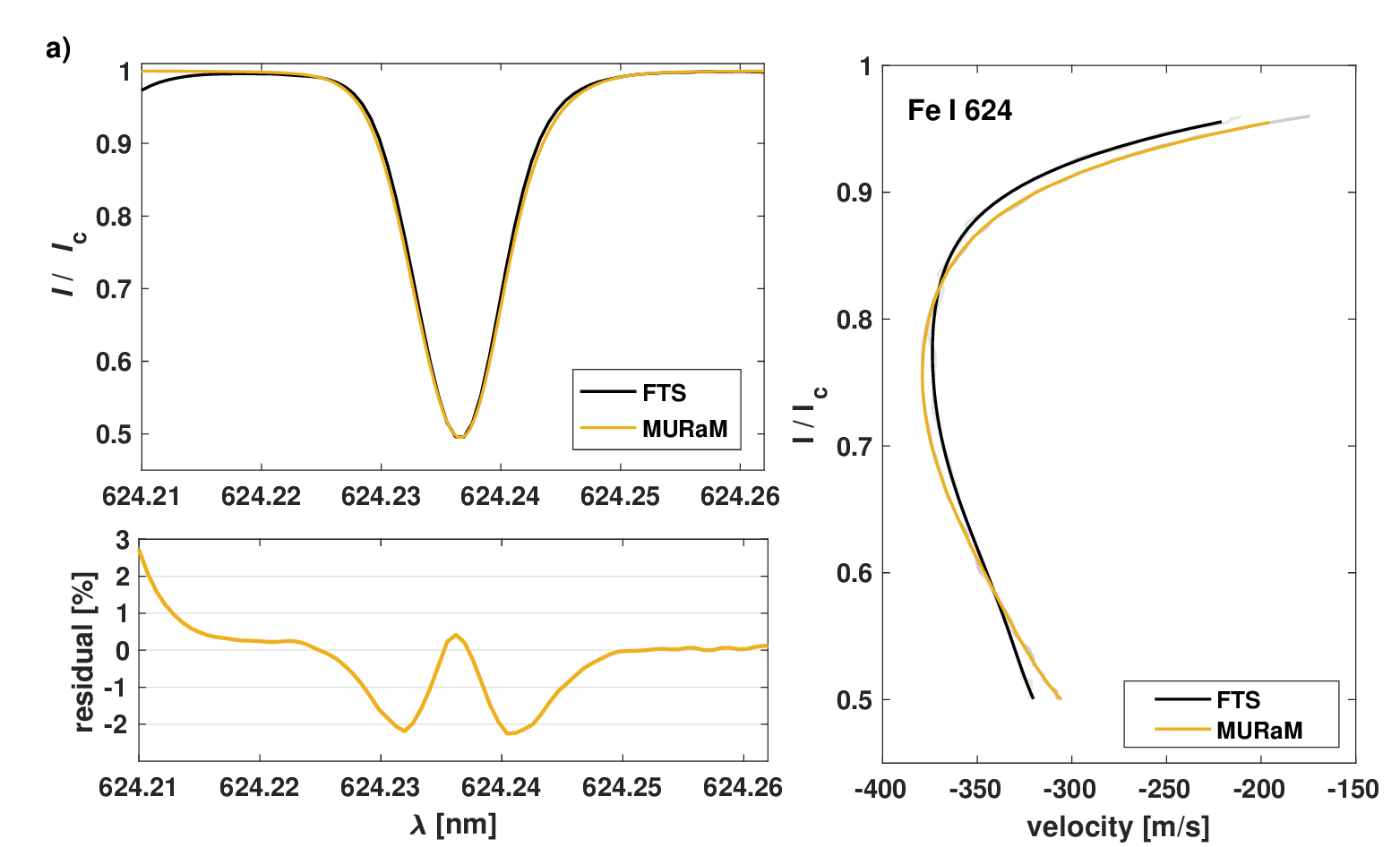}
   \includegraphics[width=\linewidth]{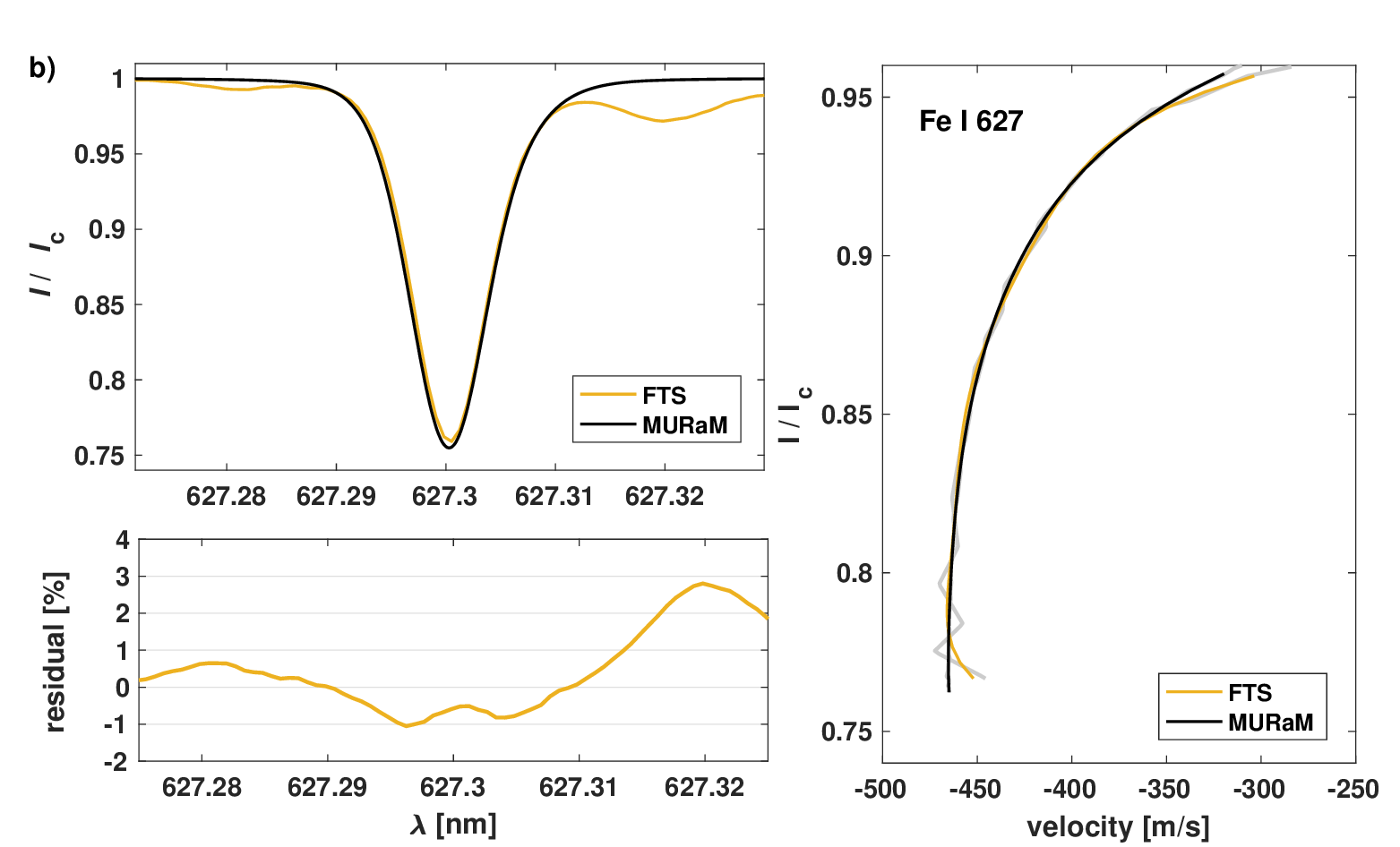}
   \includegraphics[width=\linewidth]{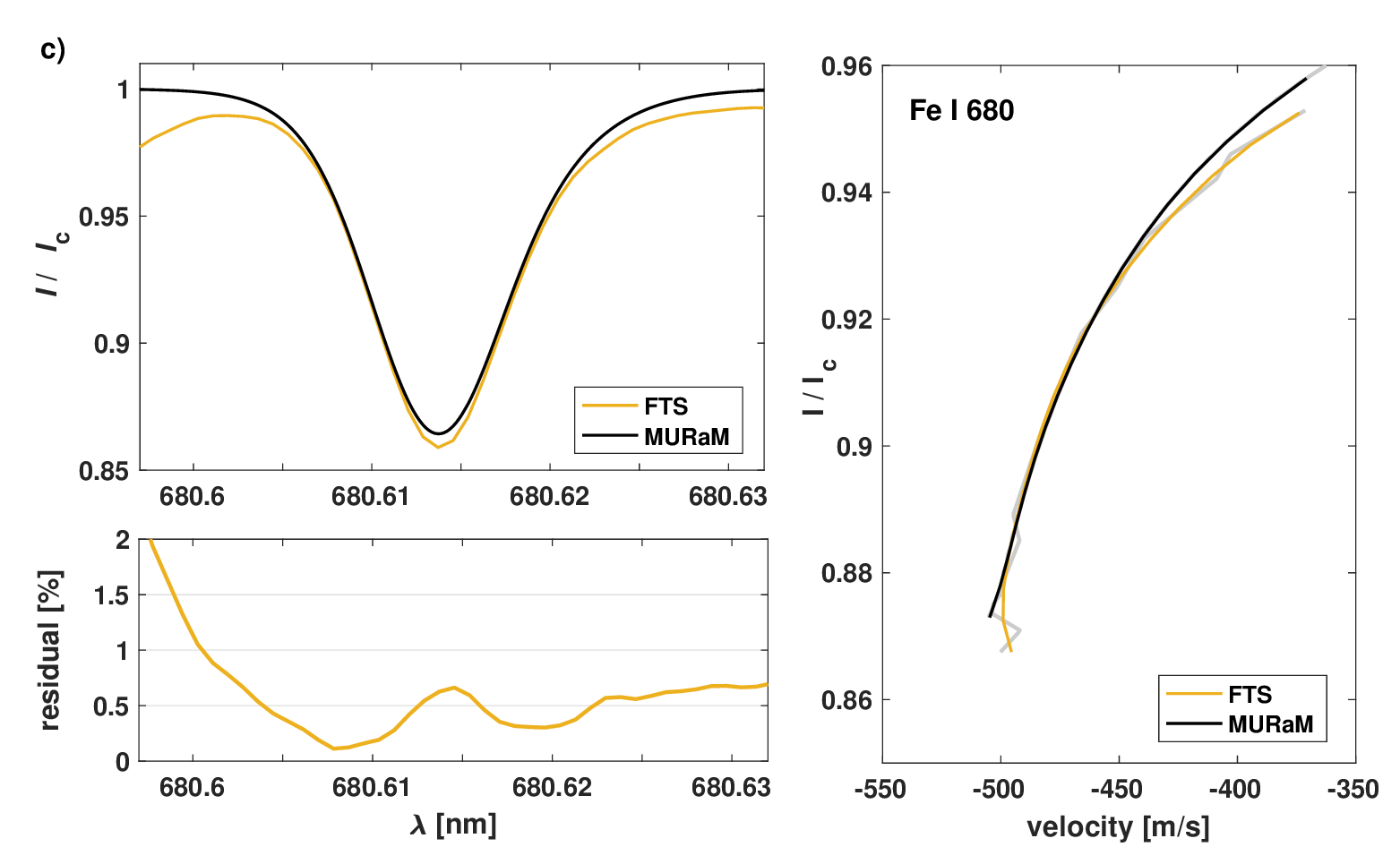}
   \includegraphics[width=\linewidth]{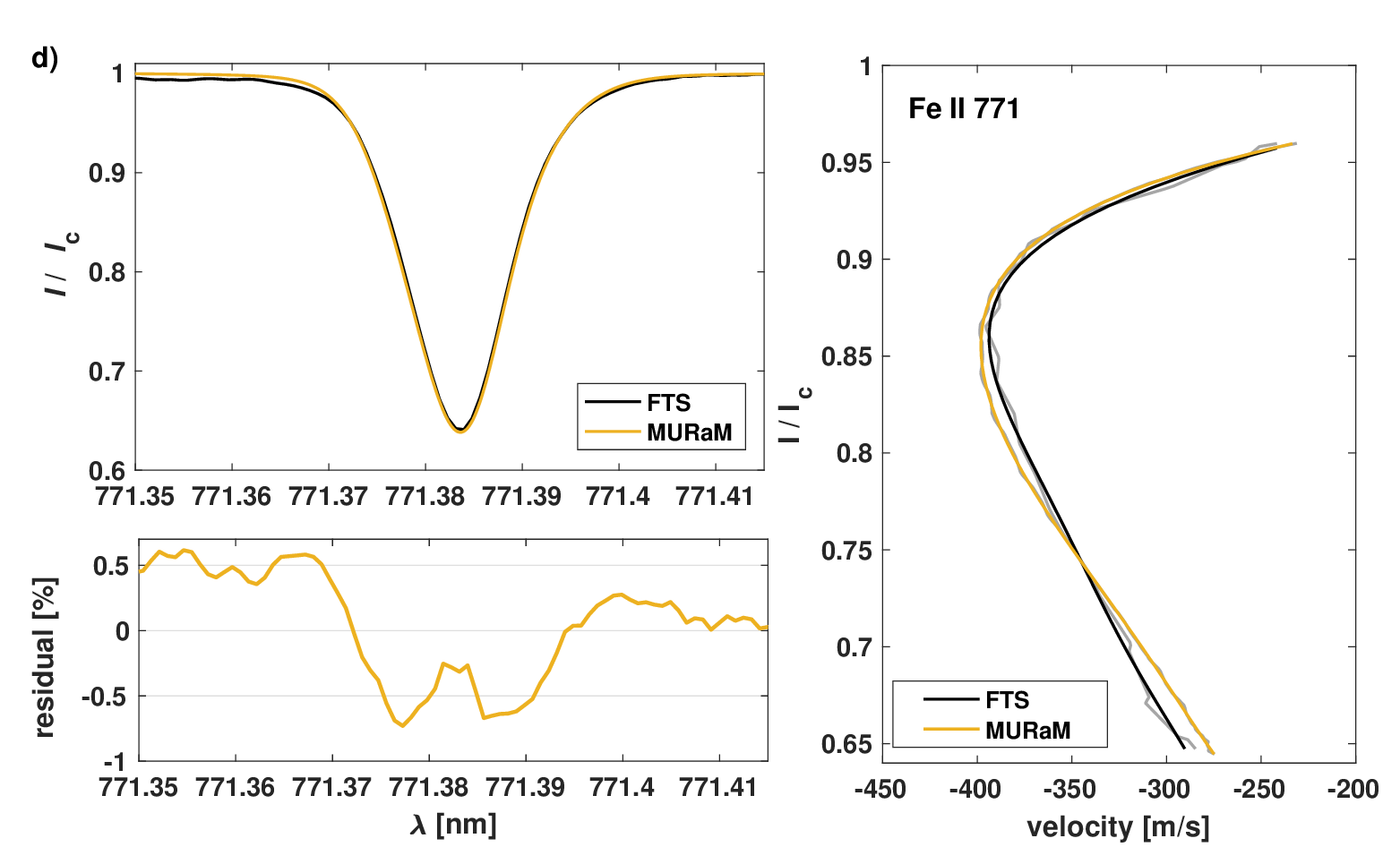}
   }
  \caption{Calculated and observed line profiles, $I/I_c$ (left top panels), and the residuals in \% of these two (left bottom panels),  and corresponding bisectors (right panels) for four lines. The bisector curves where fitted using a sixth degree polynomial (black and yellow curves) and the actual data is shown in gray. Shown are  the  Fe~I~624 line (a),  the  Fe~I~627 line (b),  the  Fe~I~680 line (c),  and Fe~II~771 (d) line. }
  \label{fig:spectral_lines_bisector}
\end{figure}

Generally, the wavelengths of spectral line center positions, $\lambda_0$, have large uncertainties \citep{nave_fe_i_linelist}. Therefore, we focus only on the spectral line shape and the resulting bisectors, but not on the position (we introduce shifts for better comparison). For the calculations, we use the atomic data from Kurucz's line list \footnote{http://kurucz.harvard.edu/linelists/linescd/}.  The spectral line intensities, $I$,  are synthesized along all rays at disk center and averaged over all cubes. Subsequently, we calculate the ratio of the averaged intensity over the averaged continuum intensity, $I/I_c$, shown in Fig.~\ref{fig:spectral_lines_bisector}. The bisectors are calculated from the averaged line profiles.

We compare the calculated disk center spectral lines and their bisectors to those from the Hamburg atlases of the solar observed spectrum recorded with the FTS at the McMath telescope \footnote{{ftp://ftp.hs.uni-hamburg.de/pub/outgoing/FTS-Atlas/}}.
The calculated $I/I_c$ for the
spectral lines shows good agreement with the observations.
The observed departures might be due to different accuracies of the atomic data or our assumption of LTE. Non-LTE effects are considerably larger in Fe I lines due to overionisation and usually lead to somewhat weaker and narrower line compared to LTE.
Moreover, our calculated bisectors agree with observations very well for all four lines, where the best agreement is achieved for the Fe~I~627 line. Since the bisector shape is very sensitive to the velocity field and horizontal temperature inhomogeneities introduced by convection, we conclude that MURaM SSD simulations accurately reproduce convection and overshoot. 

\section{Conclusions}
\label{Sec:conclusions}
We demonstrated that the most up-to-date 3D MURaM simulations that include a small-scale dynamo reproduce remarkably well the quiet-Sun observables at low spatial resolution, such as the solar spectral irradiance and the limb darkening in the continuum and in selected spectral lines. 
We showed that various improvements in the simulation setup are responsible for the agreement of the model with observations. These include a consistent EOS table, opacity table, and an updated diffusive terms, which lead to an improvement in the agreement of the computed spectral irradiance with state-of-the-art observations. Another improvement consists of the more accurate radiative energy transfer in the MURaM code produced by the increase in the number of spectral bins used for multi-grouping. This improvement leads to  an extraordinarily
 good agreement  of the computed CLV with standard measurements. This version of MURaM also reproduces line bisectors with high accuracy (which had so far not been tested using this code).  
 
We conclude that the newest MURaM setup reproduces both the temperature structure and the convective velocities of the quiet Sun. 
Having established a very good agreement between simulations and selected observables for the quiet Sun, we now have a solid justification for extending the MURaM atmospheric simulations to other stars.

\begin{acknowledgements}
This work has received funding from the European Research Council (ERC) under the European Union's Horizon 2020 research and innovation programme (grant agreement No. 715947). N.M.K. and L.G. acknowledge support from the German Aerospace Center (DLR FKZ~50OP1902).
L.G. acknowledges funding from ERC Synergy Grant WHOLE~SUN~810218. We kindly thank Qi Jin for providing us the observational data from the Solar Spectral Irradiance Monitor onboard the Fengyun-3E Satellite. We would like to thank the anonymous referee for helpful comments. 
\end{acknowledgements}

\bibliographystyle{aa} %
\bibliography{refer} %

\begin{appendix}
\section{Spectral irradiance using different 3D simulations and setups}
\label{app:irradiance}

In the following we showcase the differences  in the spectral irradiance resulting from different MURaM versions and setups.

\begin{figure}
   {\includegraphics[width=\linewidth]{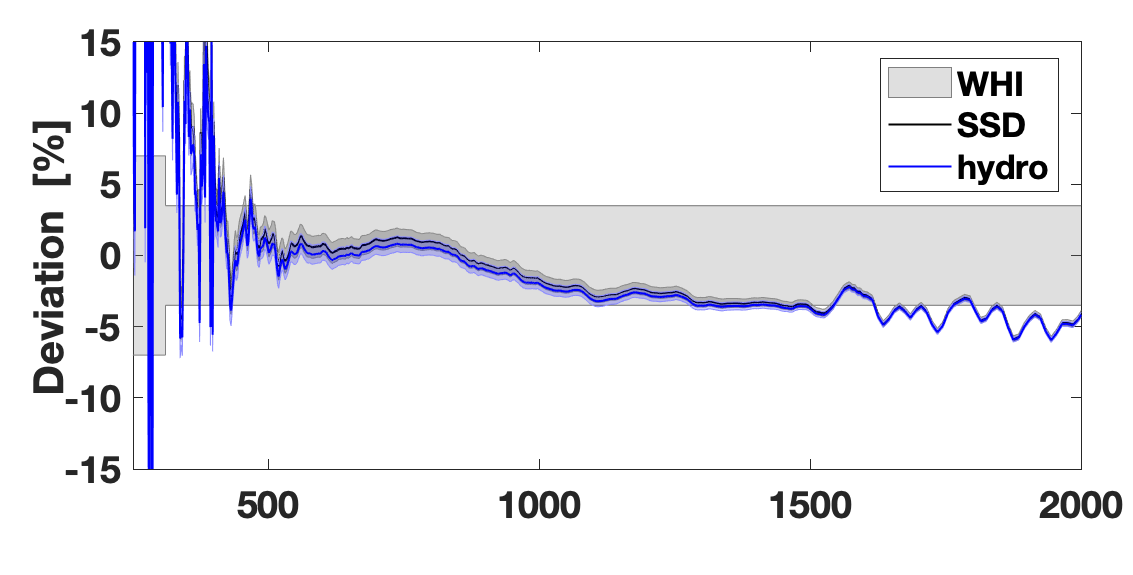}}
  \caption{Deviations (calculated as (theory - measurements)/ measurements)) of spectral irradiance between the WHI spectrum and spectral solar irradiance calculated using SSD and HD cubes.
  The light grey shaded area around the x-axis indicates the uncertainty of the WHI spectrum.
  The light blue and dark grey shaded areas around the curves representing HD and SSD simulations  indicate the error of the mean due  to the temporal fluctuations of the simulations.}
  \label{appfig:hd_ssd_irradiance}
\end{figure}

\subsection{Effect of small-scale dynamo on low-resolution spectrum}
MURaM is capable of simulating the action of a small-scale turbulent dynamo, that is the interaction between solar convection and magnetic flux \citep[see, e.g.,][and references therein]{rempel_2014}.  
Since it is computationally not feasible to simulate the whole convection zone, we use boundary conditions that successfully mimic a deeper convection zone found by \citet{Rempel_2018ApJ}. 
To demonstrate the effect of an SSD on the spectral irradiance we calculated the  deviations between the calculated to the observed spectral irradiance. Figure~\ref{appfig:hd_ssd_irradiance} shows these deviations, where the spectral irradiance for the SSD was computed using more than 300 SSD cubes and for the HD simulation more than 130 HD cubes. For these 3D simulations the radiative cooling and heating rates have been computed using 12 multi-group bins. 

The action of the SSD leads to the increase of spectral irradiance by about 1\% in the visible spectral domain but the effect of the SSD on the infrared irradiance is much smaller. We note that 
the action of the SSD strongly affects the velocity field  \citep[see bottom panel of Figure 5 in][]{Witzke_et_all_2022}. Thus, while the effect on the low-resolution spectrum is relatively small, we expect a bigger effect on the profiles of spectral lines (Mauviard et al., in prep.).

\subsection{The importance of using the consistent abundances throughout the calculations}
\begin{figure}
   {\includegraphics[width=\linewidth]{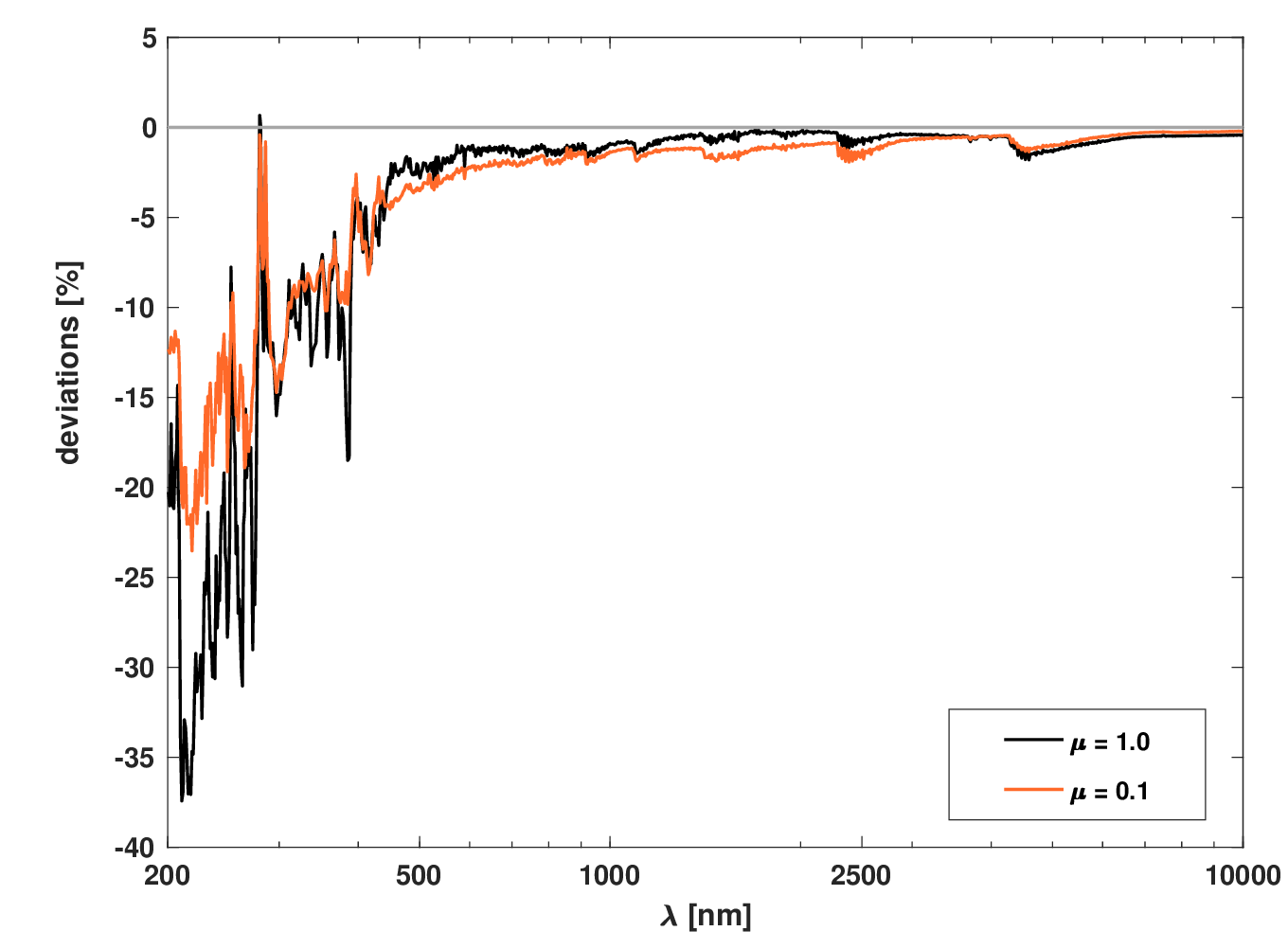}}
  \caption{Deviations of emergent intensities of one particular cube between RT calculations performed using two different sets of element compositions: first being the same as used in the 3D calculations of the underlying model (Asplund composition) and the second being the element composition from \citet{Anders_Grevesse_1989}.
  The comparison is shown for two viewing angles. }
  \label{appfig:1D_exer}
\end{figure}

Another essential feature of our calculations is a fully consistent treatment of the element composition that is used throughout all calculations:  the same element composition is used in the EOS, in the opacity tables for the 3D MURaM simulations, and in calculating spectra emergent from the 3D cubes.  
To illustrate the importance of the consistent treatment of abundances we performed the following exercise: We re-calculated the spectral irradiance from one particular cube, but instead of using the consistent Asplund composition, we used the element composition from \citet{Anders_Grevesse_1989} for calculating spectra emergent from the 3D cubes with MPS-ATLAS. Figure \ref{appfig:1D_exer} shows the deviation in the intensities when an inconsistent element composition was used for two viewing angles. It becomes evident that in the visible the deviations can be up to 10\%, decreasing towards the IR. However, in the UV deviations of more than 35\% are reached at disc centre.

\begin{figure*}
 \sidecaption
   {\includegraphics[width=12cm]{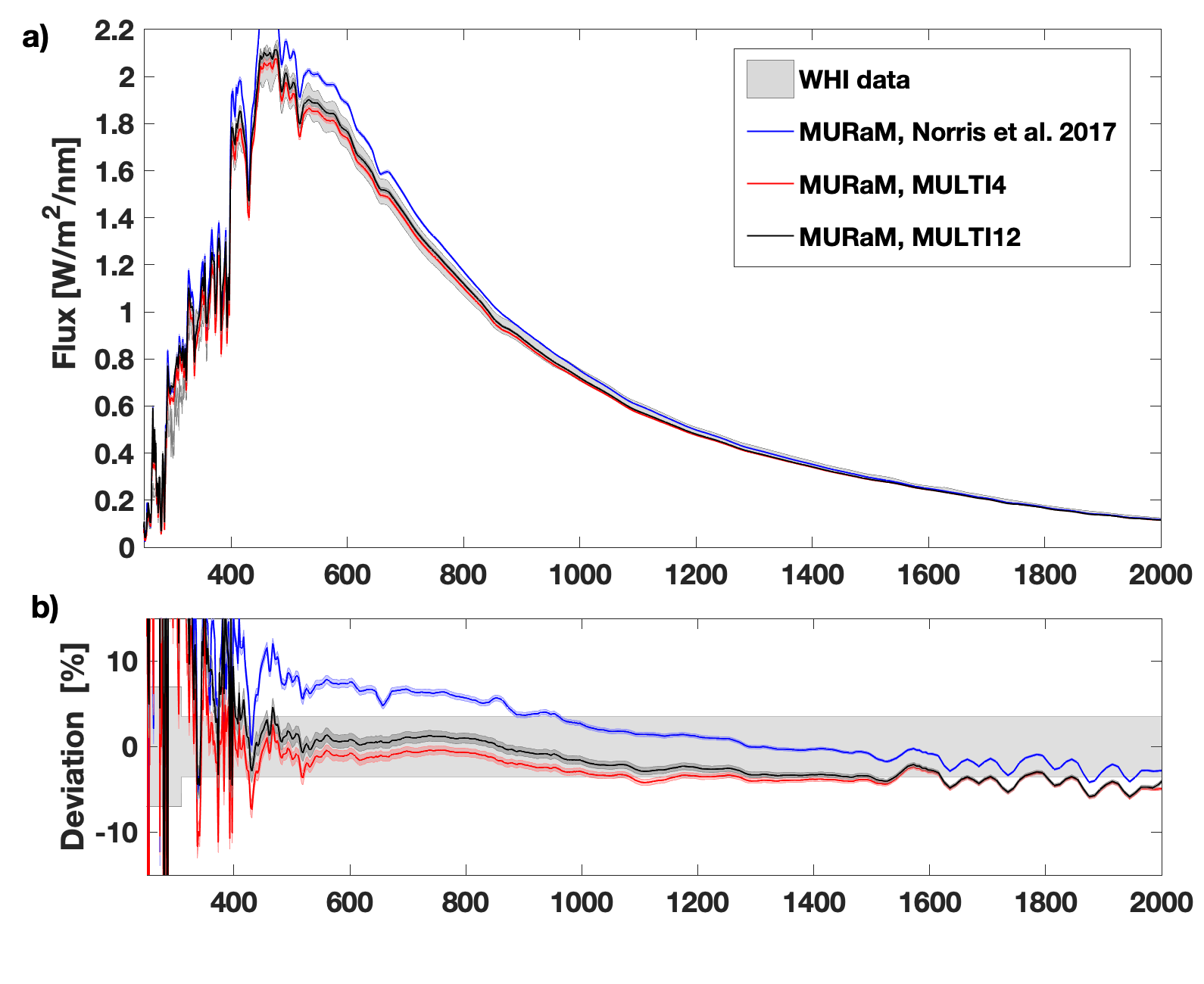}}
  \caption{Calculated irradiance spectrum using different versions and setups of the MURaM code together with the measured SIRS WHI solar irradiance spectrum. The grey shaded area indicates the measurement error of the SIRS WHI as in Fig.~\ref{fig:muram_versus_observations}. The averaged solar irradiance spectrum computed from Beeck2013 cubes are shown in blue,  where the light blue shaded area indicates one standard deviation in the temporal fluctuations. The red  and black lines show the averaged spectrum from cubes simulated with the updated MURaM code version but using different multi-group binning (MULTI4 and MULTI12). The light red and dark grey shaded area indicates one standard deviation in the temporal fluctuations of these two sets of 3D cubes.}
  \label{appfig:muram_versus_observations}
\end{figure*}

\subsection{Comparison to the Beeck2013 cubes }

Finally, we show a comparison of the spectral irradiance computed from Beeck2013 cubes  and cubes using the newest MURaM setup as in  \citet{Bhatia_2022}  performed with 4 multi-group bins  \citep{Witzke_et_all_2022} (MULTI4)  and with 12 multi-group (MULTI12) bins to obtain the radiative cooling and heating rates. 
We note that the Beeck2013 cubes were computed with the main aim of understanding the physical mechanism of near-surface convection in main sequence stars. While the Sun was modelled using realistic 3D MHD simulations, there was no goal to accurately match available observations. In particular, radiative cooling and heating rates in \citet{Beeck2013A&Afirst} have been computed using only 4 multi-group bins which is not sufficient for obtaining their precise values \citep{Andrea2023}. Furthermore, the EOS look-up tables were taken from the OPAL project \citep{OPAL_EOS_1996} and had a different element composition than the generated opacity tables.  In contrast, the simulations by  \citet{Bhatia_2022} and \citet{Witzke_et_all_2022}  have been performed to  model the Sun as accurately as possible.  The EOS was changed to the FreeEOS code \citep{Irwin_freeeos_2012} and the element composition from \citet{Asplund_2009} was used for both opacity and EOS calculations. We have also computed the cooling and heating rates with 4 multi-group bins \citep[see][]{Witzke_et_all_2022} (MULTI4)  and using 12 multi-group bins  (MULTI12) as otherwise done throughout this paper. 

To investigate the improvements in these more recent simulations, we show in Fig.~\ref{appfig:muram_versus_observations} spectral irradiance calculated with Beeck2013 cubes \citep[calculations are taken from][]{Norris_2017A&A} and spectral irradiance obtained from the MULTI4 and MULTI12 3D cubes. The comparison reveals that the Beeck2013 calculations show a slightly higher effective temperature and greater spectral irradiance in the visible spectral domain with a steeper drop towards IR than the MULTI4 and MULTI12 calculations. Focusing on the visible domain, one can see that the shape for most of the spectral line features show good agreement with the observations, but the overall flux is higher.
All in all,  both the MULTI4 and MULTI12 calculations have a much better agreement with the WHI reference spectrum. In particular, the gradient of the irradiance from the visible to the infrared matches the observation much better. This indicates that the continuum opacities and the atmospheric structure in the MULTI4 and MULTI12 simulations are more accurate.

However, the calculated spectral irradiance from MULTI12 simulations matches the WHI slightly better in the visible and near IR compared to the calculations from the MULTI4. Moreover, it shows a stronger drop towards the IR. This can be explained as follows:  In the simulation from the MULTI12 cubes the photospheric layers are heated as the RT becomes more efficient and the temperature gradient becomes less steep compared to the MULTI4 cubes with a 4 multi-group binning. This clearly manifests itself in the centre-to-limb variations presented in Appendix~\ref{app:clv}.

\section{Improvement of the CLVs}
\label{app:clv}

\begin{figure*}
  \sidecaption
   {\includegraphics[width=12cm]{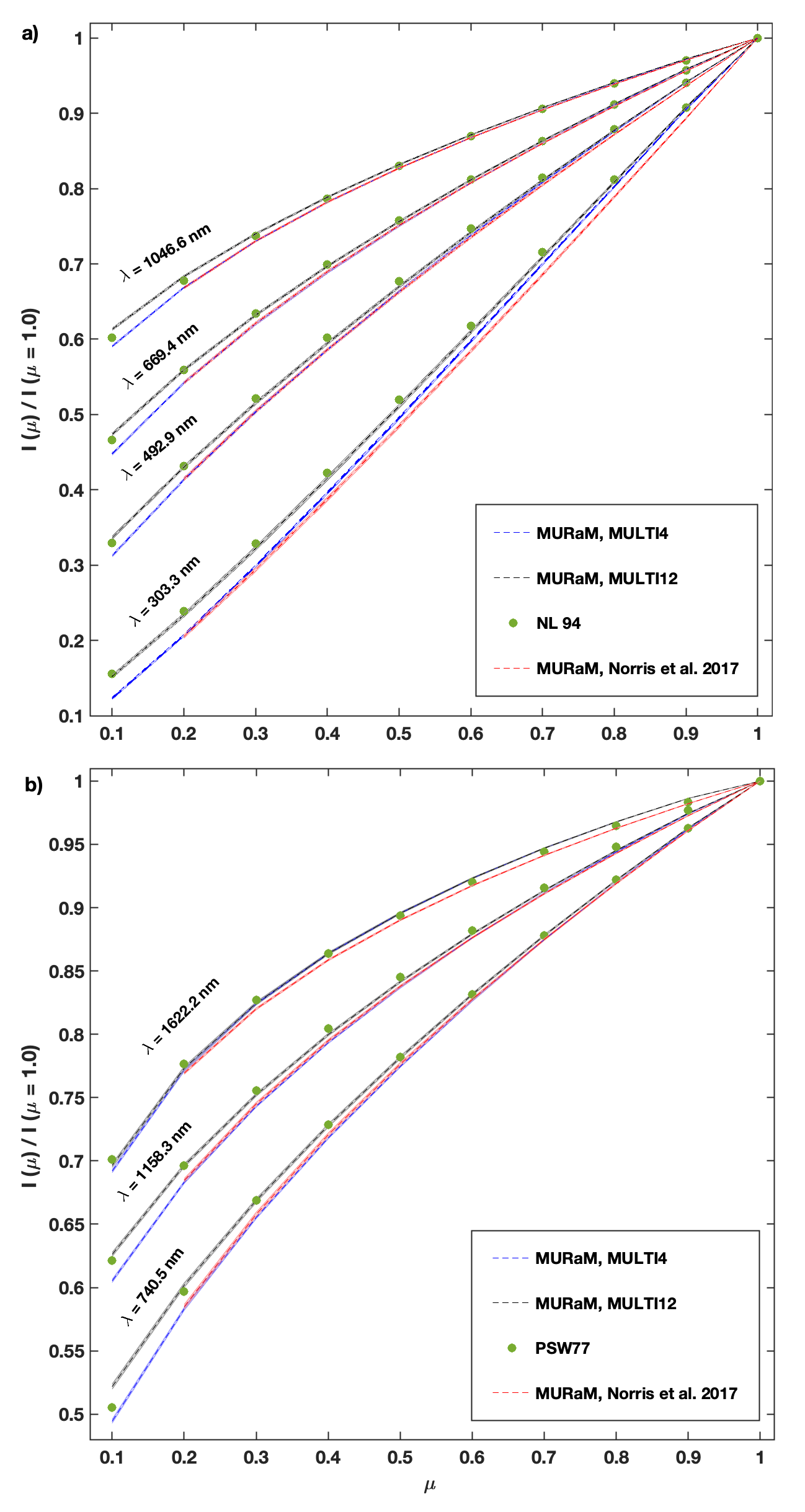}} 
  \caption{Limb darkening obtained from different MURaM cubes  ( MULTI4, MULTI12 and Beeck2013) and from solar measurements NL94 \citep{NL94} and PSW77 \citep{PSW77} at different wavelengths. A three sigma error of the mean is plotted for all MURaM results: light blue shaded for MULTI4, grey shaded for MULTI12 and light red shaded for the calculations from Beeck2013 cubes.}
  \label{appfig:02-ld-comparison}
\end{figure*}

The CLVs are more sensitive to the temperature gradient in the photosphere than the spectral irradiance. Figure \ref{appfig:02-ld-comparison} displays the CLVs from the Beeck2013 simulations \citep[calculations are taken from][]{Norris_2017A&A}, from the MULTI4 and MULTI12 simulations by \citet{Witzke_et_all_2022}.

One can see that the limb darkening computed from the Beeck2013 simulations and from MULTI4 cubes is much steeper than the observed CLVs. The MULTI12 simulations  lead to a smaller drop of intensity towards the limb and, thus, a much better agreement with observations.   
This is because in the MULTI12 simulations the photosphere is further heated due to the more efficient RT. Such a heating leads to a shallower temperature gradient and a more accurate CLV. Since the he CLV from the MULTI12  cubes leads to an excellent agreement with observations, we conclude that 12 multi-group bins are sufficient to accurately model the RT in 3D simulations at least for continuum wavelengths.  

The difference between MULTI4 and MULTI12 is smallest at  1622.2~nm, which  is close to the H- opacity minimum, where the CLVs from  MULTI4 and MULTI12 cube computations almost coincide. This is because the 
intensity at this wavelength is formed at the deepest photospheric layers, where the energy is transported mainly by convective motions and, thus, a more efficient RT does not have a big impact on the temperature structure.

\end{appendix}

\end{document}